\documentclass[useAMS,usenatbib]{mn2e}
\usepackage{graphicx}
\usepackage{amssymb}
\usepackage{enumerate}

\footnotesize
\newdimen\digitwidth    
\setbox0=\hbox{\rm0}
\digitwidth=\wd0
\catcode`!=\active
\def!{\kern\digitwidth}
\normalsize

\newcommand{\psr}{PSR~J1614$-$2230}

\title[\psr: Evidence for a NS born massive]
      {Formation of millisecond pulsars with {CO} white dwarf companions  
      -- I. \psr: Evidence for a neutron star born massive}

\author[Tauris, Langer \& Kramer]{
  T.~M.~Tauris,$^1$$^,$$^2$ N.~Langer$^1$$^,$$^3$ and M.~Kramer$^2$$^,$$^4$
  \\
  $^{1}$Argelander-Insitut f\"ur Astronomie, Universit\"at Bonn, Auf
  dem H\"ugel 71, 53121 Bonn, Germany 
  \\
  $^{2}$Max-Planck-Institut f\"ur Radioastronomie, Auf dem H\"ugel 69,
  53121 Bonn, Germany
  \\
  $^{3}$Astronomical Institute, Utrecht University, Princetonplein 5, 
  3584 CC, Utrecht, The Netherlands
  \\
  $^{4}$Jodrell Bank Centre for Astrophysics, University of Manchester,
  Oxford Rd, Manchester M13 9PL, UK 
  \\
  }
\date{Drafted 12 May 2011.}
\begin{document}

\maketitle

\begin{abstract} 
The recent discovery of a $2\,M_{\odot}$ binary millisecond pulsar \citep{dpr+10} has not only important
consequences for the equation-of-state of nuclear matter at high densities but also raises the
interesting question if the neutron star \psr\ was born massive. The answer
is vital for understanding neutron star formation in core collapse supernovae. Furthermore, this system
raises interesting issues about the nature of the progenitor binary and how it evolved during its
mass exchanging X-ray phase.
In this paper we discuss the progenitor evolution of \psr.
We have performed detailed stellar evolution modelling of intermediate-mass {X}-ray binaries undergoing Case~A Roche-lobe overflow (RLO)
and applied an analytic parameterization for calculating the outcome of either a common envelope evolution or the highly super-Eddington isotropic re-emission mode.
We find two viable possibilities for the formation of
the \psr\ system: either it contained a $2.2-2.6\,M_{\odot}$ giant donor star and evolved through 
a common envelope and spiral-in phase
or, more likely, it descended from a close binary system with 
a $4.0-5.0\,M_{\odot}$ main sequence donor star via Case~A RLO. We conclude that the
neutron star must have been born with a mass of $\sim\!\!1.95\,M_{\odot}$ or $1.7\pm0.15\,M_{\odot}$, respectively -- 
which significantly exceeds neutron star birth masses in previously discovered radio pulsar systems.
Based on the expected neutron star birth masses from considerations of stellar evolution
and explosion models, we find that the progenitor star of \psr\ is likely to have been more massive than $20\,M_{\odot}$.

\end{abstract}



\section{Introduction}
\label{sec:Intro}
Neutron stars are formed as compact remnants of massive stars ($10-30\,M_{\odot}$) which explode
in supernovae at the end of their stellar life \citep{whw02,hfw+03}.
In order to better understand the mechanisms of the electron capture and core collapse supernovae
knowledge of the distribution of birth masses of neutron stars is vital.
However, in order to weigh a neutron star it must be a member of a binary system.
This introduces an uncertainty in determining the original birth mass of the neutron star
since these neutron stars are often observed in {X}-ray binaries or, at a later stage, as recycled pulsars
and hence {\it after} they have undergone a phase of mass accretion from their companion star.
The most precisely measured masses of neutron stars are obtained in double neutron star systems
via general relativistic effects. The related post-Keplerian parameters include periastron advance, 
redshift/time~dilation, orbital period derivative and Shapiro delay \citep[e.g.][]{Wil09}.  
Shapiro delays of radio signals from pulsars \citep[][]{sac+98} have the advantage
of being measurable also in low eccentricity systems if the orbital inclination is such that
the pulses passes in the vicinity of its companion.
This method yields the opportunity to
weigh both neutron stars accurately -- and hence also determine the mass of the
last formed neutron star which has not accreted any material.
So far, such measurements have revealed that even the most massive of these neutron stars (the non-recycled pulsars)
do not exceed a mass of $1.39\,M_{\odot}$ \citep{tc99,spr10}.
There is, however, some evidence from neutron stars in {X}-ray binaries, e.g. Vela~{X-1}, that 
suggests neutron stars can be born more massive than this value.

Binary millisecond pulsars are known to be key sources of research in fundamental physics. They host the densest
matter in the observable Universe and possess very rapid spins as well as relativistic magnetospheres with 
outflowing plasma winds.
Being ultra stable clocks they also allow for
unprecedented tests of gravitational theories in the strong-field regime \citep{kw09}. Equally important, however,
binary millisecond pulsars represent the end point of stellar evolution, and their observed orbital
and stellar properties are fossil records of their evolutionary history. Thus one
can use binary pulsar systems as key probes of stellar astrophysics.\\
Recent Shapiro delay measurements of \psr\ \citep{dpr+10} allowed a precise mass determination
of this record high-mass pulsar (neutron star) and its white dwarf companion.
Characteristic parameters of the system are shown in Table~\ref{table:param}.
It is well established that the neutron star in binary millisecond pulsar systems forms first, descending from the
initially more massive of the two binary stellar components. The neutron star is subsequently spun-up to a high spin frequency
via accretion of mass and angular momentum once the secondary star evolves
\citep{acrs82,rs82,bv91}. In this recycling phase the
system is observable as a low-mass X-ray binary \citep[e.g.][]{nag89} and towards the end of this phase
as an {X}-ray millisecond pulsar \citep{wv98,asr+09}.
Although this formation scenario is now commonly accepted
many aspects of the mass-transfer process and the accretion physics (e.g. the accretion efficiency
and the details of non-conservative evolution) are
still not well understood \citep{lv06}.

In this paper we investigate the progenitor evolution of \psr. We are mainly focusing on the important 
{X}-ray binary phase starting 
from the point where the neutron star has already formed. However, we shall also briefly
outline the previous evolution from the zero-age main sequence (ZAMS)~binary until this stage since
this evolution is important for the birth mass of the neutron star.
In Section~\ref{sec:masstransfer} we discuss the three different possibilities for mass transfer toward a neutron star from an
intermediate-mass star of $2.2 - 5.0\,M_{\odot}$, for the Roche-lobe overflow (RLO) Cases A, B and C. 
The evolution of the original ZAMS binary until the X-ray phase is briefly discussed in Section~\ref{sec:progenitor}.
In Section~\ref{sec:discussion} we compare our results with the outcome of the independent work by
\citet{lrp+11} and also discuss our results in a broader context in relation to neutron star birth masses
predicted by stellar evolution and supernova explosions.
Our conclusions are given in Section~\ref{sec:conclusions}.\\ 
In Paper~II \citep{tlk11b} we continue the
discussion of \psr\ in view of general aspects of accretion onto neutron stars during the recycling process of millisecond pulsars.


\begin{table}
\center
\caption{Physical parameters of the binary millisecond pulsar \psr\ 
         \citep[data taken from ][]{dpr+10}.}
\begin{tabular}{lr}
\hline {Parameter} & {value} \\ 
\hline 
\noalign{\smallskip} 
            Pulsar mass & $1.97\pm 0.04\,M_{\odot}$ \\
            White dwarf mass & $0.500\pm 0.006\,M_{\odot}$ \\
            Orbital period & $8.6866194196(2)\;\rm{days}$ \\
            Projected pulsar semimajor axis & $11.2911975\;\rm{light~sec}$ \\
            Orbital eccentricity & $1.30\pm 0.04 \times 10^{-6}$ \\
            Inclination angle & $89.17\pm 0.02\;\rm{deg.}$ \\
            Dispersion-derived distance & $1.2\;\rm{kpc}$ \\
            Pulsar spin period & $3.1508076534271\;\rm{ms}$ \\
            Period derivative & $9.6216\times 10^{-21}$ \\
\noalign{\smallskip} 
\hline
\end{tabular}
\label{table:param}
\end{table}


\section{Mass transfer in {X}-ray binaries}
\label{sec:masstransfer}

\begin{figure*}
\begin{center}
  \mbox{\includegraphics[width=0.60\textwidth, angle=-90]{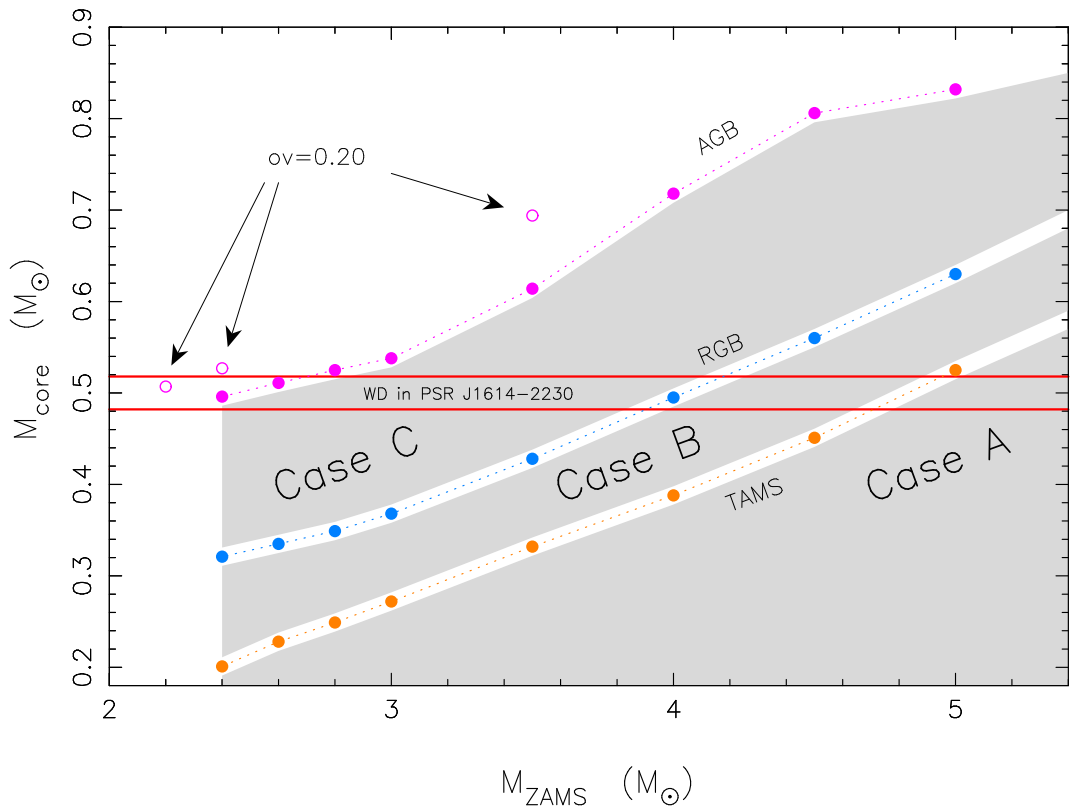}}
  \caption[]{
    Stellar core mass as a function of the initial ZAMS mass, calculated at
    different evolutionary epochs. Mass transfer initiated either before the TAMS, the tip of the RGB or the AGB,
    corresponds to RLO Case~A, Case~B and Case~C, respectively. Filled circles represent models
    without convective core-overshooting. The open circles show a few examples of core masses
    at the tip of the AGB assuming a convective core-overshooting parameter of $\delta _{\rm OV}=0.20$.
    The horizontal red lines indicate the measured mass interval (within 3-$\sigma$ error bars) of the white dwarf 
    in \psr. This white dwarf descends from the core of the donor star in the {X}-ray binary.
    }
\label{fig:coremass}
\end{center}
\end{figure*}

Consider a close interacting binary system which consists of a non-degenerate donor star and a compact object, in our case a neutron star.
If the orbital separation is small enough the (evolved) non-degenerate star fills its inner common equipotential
surface (Roche-lobe) and becomes a donor star for a subsequent epoch of mass transfer toward the, now, accreting neutron star.
In this phase the system is observed as an {X}-ray binary.
When the donor star fills its Roche-lobe it is perturbed by removal of mass and it falls out of hydrostatic
and thermal equilibrium. In the process of re-establishing equilibrium the star will either grow or shrink
-- depending on the properties of its envelope layers as discussed below -- first on a dynamical (adiabatic)
timescale and subsequently on a slower thermal timescale. However,
any exchange and loss of mass in such an {X}-ray binary system will also lead to alterations of the orbital dynamics,
via modifications in the orbital angular momentum,
and hence changes in the size of the critical Roche-lobe radius of the donor star. The stability of the
mass-transfer process therefore depends on how these two radii evolve (i.e. the radius of the star and the Roche-lobe radius).
The various possible
modes of mass exchange and loss
include, for example, direct fast wind mass loss, Roche-lobe overflow, with or without isotropic re-emission,
and common envelope evolution \citep[e.g.][and references therein]{vdh94a, spv97}.
The RLO mass transfer can be initiated while the donor star is still on the main sequence (Case~A RLO), during hydrogen
shell burning (Case~B RLO) or during helium shell burning (Case~C RLO). 
The corresponding evolutionary timescales for these different cases will in general proceed on
a nuclear, thermal or dynamical timescale, respectively, or a combination thereof.
We now investigate each of these three cases with the
aim of reproducing the parameters of \psr.

\subsection{Case~C RLO - dynamical unstable mass transfer}\label{subsec:CaseC}
Donor stars in systems with wide orbits ($P_{\rm orb}\simeq 10^2-10^3$~days) prior to the mass-transfer phase
develop a deep convective envelope as they become giant stars before filling their Roche-lobe.
The response to mass loss
for these stars with outer layers of constant low entropy and negative adiabatic mass-radius exponents
($\xi=\partial \ln R/\partial \ln M <0$) is therefore expansion
which causes the stars to overfill their Roche-lobes even more.
To exacerbate this problem, binaries also shrink in size if mass transfer occurs from a
donor star somewhat more massive 
than the accreting neutron star. This causes further
overfilling of the donor star Roche-lobe resulting in enhanced mass loss etc.
This situation is clearly a vicious circle that leads to a runaway mass transfer and the formation of a
contact binary with a common envelope (CE) followed by a spiral-in phase, e.g. \citet{pac76}, \citet{il93}.

A simple estimate of the reduction of the orbit can be found by
equating the binding energy of the envelope of the AGB giant donor to the
difference in orbital energy before and after the CE-phase. The idea is that the motion of the
neutron star, once captured in the CE, results in friction and thus dissipation of
orbital energy which can be used to expel the CE. Following the formalism
of \citet{web84} and \cite{dek90}, the binding energy of the envelope at the onset of
RLO mass transfer can be written as: $-G M_2 M_{\rm env} /
(\lambda\,R_2)$, where $M_2$ is the mass of the donor star, with envelope mass $M_{\rm env}$,
and $R_{\rm 2}=R_{\rm L}$ is the Roche-lobe radius of the donor star at the onset of the CE-phase.
This radius is often calculated in terms of its dimensionless Roche-lobe radius, $r_{\rm L}$ \citep{egg83} such that
$R_2 \simeq R_{\rm L} = a_0 \cdot r_{\rm L}$, where
$a_0$ is the initial orbital separation.\\

The total binding energy of the envelope
includes both the negative gravitational binding energy and the positive thermal energy. Besides from
the thermal energy of a simple perfect gas, the latter term also includes the energy of radiation,
terms due to ionization of {H} and {He} and dissociation of {H$_2$}, as well as the
contribution from the Fermi energy of the degenerate electrons \citep{hpe94, hpe95}.
The value of the $\lambda$-parameter can thus be calculated
from stellar structure models \citep{dt00, dt01, tad01, xl10a, xl10b, lvk11, iva11}.
Given the radius of the
donor star and the $\lambda$-parameter enables one to estimate the change in orbital
separation as a result of the neutron star spiral-in and ejection of the envelope.
Let $\eta_{\rm ce}$ describe the efficiency
of ejecting the envelope via drag forces, i.e.
of converting orbital energy ($E_{\rm orb}=-G M_2 M_{\rm NS}/2\,a$) into the kinetic energy that provides the outward
motion of the envelope: $E_{\rm env} \equiv \eta_{\rm ce} \,\, \Delta
E_{\rm orb}$ and one finds the well-known expression for the ratio of the change in orbital separation:
\begin{eqnarray}
    \frac {a}{a_0} & = &
    \frac {M_\mathrm{core} M_\mathrm{NS}} {M_2}
    \frac {1} {M_\mathrm{NS} + 2 M_\mathrm{env}/
          (\eta_\mathrm{ce}\lambda r_\mathrm{L})}
    \label{aa0_CE_webbink}
\end{eqnarray}
where $M_{\rm core} = M_2 - M_{\rm env}$ is the core mass of the evolved donor star
(essentially the mass of the white dwarf to be formed);
$M_{\rm NS}$ is the mass of the neutron star and $a$ is the final orbital separation after the
CE-phase. Strictly speaking, when considering the energy budget the "effective efficiency parameter" should also
include the excess energy of the ejected matter at infinity -- although this effect is probably small.
Recent work \citep{zsgn10,dpm+11} suggests that the efficiency parameter is of the order
$30\,\%$, i.e. $\eta_{\rm ce}\simeq 0.3$, although its uncertainty is large. The value may not be universal
and could, for example, depend on the stellar mass ratio in a given binary.\\
During the very short spiral-in and ejection phase of a common envelope evolution ($\sim\!10^3$ yr) it is a
good approximation to assume that the neutron star does not accrete any significant amount of matter
given that its accretion is limited by the Eddington luminosity corresponding to
a maximum accretion rate of $\sim\!10^{-8}\,M_{\odot}\,{\rm yr}^{-1}$, depending on
the exact chemical composition of the accreted material and the geometry of its flow.

The possibility of hypercritical accretion onto the neutron star during the spiral-in phase
has been suggested to lead to significant mass increase and possible collapse of the
neutron star into a black hole \citep{che93, bro95, fbh96}. However, there is 
observational evidence that, at least in some cases, this is not the case. The recent determination of the low
pulsar mass in PSR~J1802$-$2124 of $1.24\pm 0.11\,M_{\odot}$ \citep{fsk+10} clearly demonstrates that
this $12.6~$millisecond recycled pulsar did not accrete any significant amount of matter.
This pulsar is in a tight binary with an orbital period of only 16.8~hours
and it has a carbon-oxygen white dwarf ({CO}~WD) companion of mass $0.78\pm 0.04\,M_{\odot}$.
With such a small orbital period combined with a massive white dwarf there seems to be no 
doubt that this system evolved through a common envelope and spiral-in phase and
the low pulsar mass reveals
that very little mass has been accumulated by the neutron star during this phase.
However, the recycling of this pulsar does require some $10^{-2}\,M_{\odot}$ of accreted material 
(see further discussion in Paper~II). While in principle,
PSR~J1802$-$2124 could have formed via accretion-induced collapse of an {O-Ne-Mg} WD,
so far there are no models which suggest this.

Before we proceed to discuss the case of \psr\ let us introduce a new parameterization
for calculating the outcome of a common envelope evolution.
We define a mass ratio parameter $k\equiv q_0/q$ where $q_0$ and $q$ represent the initial and final
ratio, respectively, of the donor star mass to the neutron star mass. Assuming the neutron star
mass to be constant during the CE-phase we can also write $k = M_2/M_{\rm core}$.
The value for $k$ is thus
the mass of the donor star, in units of its core mass, at the onset of the RLO.
This allows for a convenient rewriting of Eq.~(\ref{aa0_CE_webbink}):
\begin{equation}
  \frac{a}{a_0} = \frac{k^{-1}}{1+2q(k-1)/\eta_{\rm ce}\lambda r_L}
                = \left[ k+ 2q_0(k-1)/\eta_{\rm ce}\lambda r_L \right] ^{-1}
  \label{aa0_ce}
\end{equation}
The post-CE value for the mass ratio $q=M_{\rm WD}/M_{\rm NS}\simeq0.25$ is the present value in
\psr, which is directly determined from measurements (see Table~\ref{table:param}).
Hence we have $k = M_2/M_{\rm WD}$.
Taking $M_{\rm WD}=0.50\,M_{\odot}$ as the core mass we must first determine the value
of $M_2$ (and thus $k$) from stellar evolution calculations. To this purpose we
used a detailed one-dimensional hydrodynamic stellar evolution code. This code has been described in detail
e.g. in \citet{hlw00}.
Using solar chemical abundances ($Z=0.02$) and a mixing-length parameter of $\alpha=l/H_{\rm p}=1.5$ \citep{lan91} 
we find $2.4 \leq M_2/M_{\odot} \leq 2.6$, see Fig.~\ref{fig:coremass}, if we disregard
core convective overshooting. Including a core convective overshooting parameter
of $\delta _{\rm OV}=0.20$ \citep{cla07} allows for donor masses as low as
$2.2\,M_{\odot}$ to produce a final WD mass of $0.50\,M_{\odot}$.
Hence, $4.4\leq k\leq 5.2$ and we can now
use Eq.~(\ref{aa0_ce}) to find the pre-CE orbital separation, $a_0$ and hence the radius of the
Roche-lobe filling donor star.

\begin{figure}
\begin{center}
  \mbox{\includegraphics[width=0.35\textwidth, angle=-90]{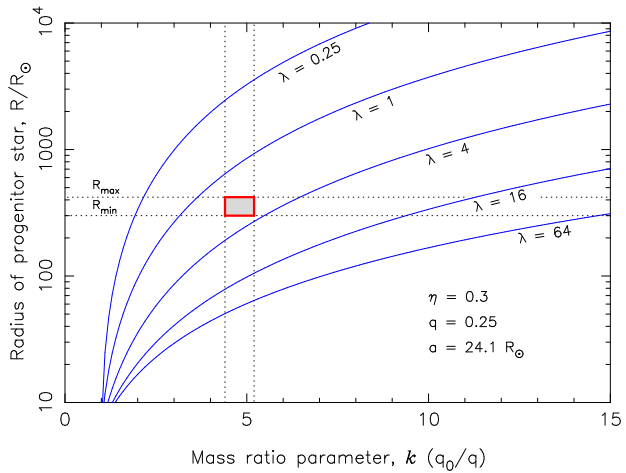}}
  \caption[]{
    Constraints on stellar parameters assuming common envelope evolution
    in the \psr\ progenitor system. The shaded area in the red box indicates the allowed parameter space
    for the radius of the progenitor star at onset of RLO and the
    mass ratio parameter, $k$.
    The various curves correspond to different $\lambda$-values of the binding
    energy of the envelope (see text).
    }
\label{fig:caseC}
\end{center}
\end{figure}

In Fig.~\ref{fig:caseC} we demonstrate that indeed \psr\ could have evolved from a CE.
The shaded rectangular area shows the parameter space of solutions. The $k$-values are constrained by
the initial donor mass, $M_2$ which in turn is constrained by the observed white dwarf mass, $M_{\rm WD}$.
The upper limit for the radius of the donor star at the onset of the RLO is simply its maximum possible radius on the AGB, $R_{\rm max}$.
We notice from the curves in the figure that only $\lambda$-values larger than about 2 are in agreement with this constraint
(i.e. of having a donor radius less than $R_{\rm max}$).
The reason for this is the relatively wide orbit of \psr\ with an orbital separation of $24.1\,R_{\odot}$. Hence,
only a modest amount of orbital energy was released during spiral-in -- almost independent of the pre-CE separation, $a_0$ since usually $a \ll a_0$ --
and therefore the binding energy of the donor star envelope cannot have been too large for a successful envelope ejection
($E_{\rm bind}\propto \lambda ^{-1}$).
The lower limit of the progenitor star radius at $\sim\!300\,R_{\odot}$ is therefore determined by exactly this requirement of having an envelope
with small enough binding energy (in this case corresponding to $\lambda \ge 2$) such that
it can be successfully ejected during the spiral-in phase. (For a graphical example of a slightly more massive donor star of $3\,M_{\odot}$, 
see Fig.~\ref{fig:coremass} in Dewi~\&~Tauris,~2000).
If the donor radius is smaller at the time of the onset of the CE then its $\lambda$-value is too small (i.e. its envelope binding
energy is too large, on an absolute scale, to allow ejection from the available orbital energy release).
The outcome is a merger event -- possibly leading to a Thorne-$\dot{\rm{Z}}$ytkow object \citep{tz77}.
A similar fate is expected for donor stars of {\it late} Case~B RLO. These stars also possess a deep convective envelope, resulting in a CE evolution. 
However, they are less evolved than stars on the AGB and have much smaller $\lambda$-values
and hence more tightly bound envelopes which strengthens the case for a merger.\\
To summarize, based on the orbital dynamics and the masses of the two stellar components, Case~C RLO (leading to a CE and spiral-in) is possible to have occurred in \psr.
This would have the implication that the neutron star was born massive
with a mass close to its presently observed  mass of $1.97\,M_{\odot}$. However, see further discussion in Sections~\ref{sec:progenitor} and \ref{sec:discussion},
and also in paper~II.

\subsubsection{Case~BB RLO following a common envelope?}\label{subsubsec:CaseBB}
One might ask whether the $0.50\,M_{\odot}$ {CO}~WD in \psr\ could have formed via Case~BB RLO from a $\sim\!1\,M_{\odot}$ helium star 
which had previously lost its hydrogen envelope in a common envelope phase. In this case the neutron star could have accreted significantly 
and needed not be so massive at birth (only about $1.5\,M_{\odot}$ if it accreted $0.5\,M_{\odot}$ subsequently). 
We have tested this hypothesis and find that it is
not a possible formation channel. The reason is that there is not enough orbital energy available (by an order of magnitude) to eject
the envelope of the WD progenitor already on the red giant branch (RGB). Assume the progenitor of the naked $1\,M_{\odot}$ helium core 
was a $5-7\,M_{\odot}$ star (the exact mass depends e.g. on the assumed amount of core convective overshooting). 
For such a star on the RGB we find that the (absolute) binding energy of the envelope is $1.3-2.2\times 10 ^{48}\,{\rm erg}$ -- according to
the formula for the binding energy in Section~\ref{subsec:CaseC} and using Fig.~1 in \citet{dt00} for estimating 
the largest possible $\lambda$-values to be $0.4-0.6$. The Case~BB RLO (from the less massive helium star to the more massive neutron star) 
would have widened the orbit to its currently observed orbital period of 8.69~days. Assuming a conservative
mass transfer the initial orbital separation would have been about $10.7\,R_{\odot}$.
This value is then equivalent to the
post-CE orbital separation corresponding to an orbital energy of $-2.7\times 10^{47}\,{\rm erg}$. Hence, even if 
all available orbital energy was released from an initial separation at infinity there would be far too little energy
to eject the envelope of the RGB star ($\Delta E_{\rm orb} \ll E_{\rm env}$). 
Only on the AGB is the envelope loose enough to allow ejection from spiral-in
of the neutron star. But in this case the core has already burned its helium into carbon and oxygen and then we are
back where we began in Section~\ref{subsec:CaseC}.

\subsection{Early Case~B RLO - thermal timescale mass transfer}\label{subsec:CaseB}
If a $3-5\,M_{\odot}$ donor star fills its Roche-lobe shortly after leaving the main sequence ({\it early} case~B) its envelope
is still radiative and the binary may survive thermal timescale mass transfer. This was shown
a decade ago in three independent papers: \citet{kr99} and \citet{pr00} studied the
formation and evolution of Cyg~{X}-2 and \citet{tvs00} investigated the
formation of binary millisecond pulsars with a CO~WD companion.
Although \citet{tvs00} and \citet{prp02}
have demonstrated that one can form systems with
a $0.50\,M_{\odot}$ CO~WD and $P_{\rm orb}=8.7\;\rm{days}$ (as observed in \psr) they 
both assumed in their calculations
an initial canonical neutron star mass of $1.30-1.40\,M_{\odot}$, which does not apply in this scenario since the neutron star only accretes
a few $0.01\,M_{\odot}$ during the Case~B rapid mass-transfer phase, thus disqualifying the neutron star from reaching its
present mass of $1.97\,M_{\odot}$.
If the initial mass ratio between the donor star and the neutron star is of the order $q_0 \simeq 2-3$ the orbit shrinks
significantly in response to mass loss, as mentioned previously. This leads to highly super-Eddington mass-transfer rates and hence
we can apply the isotropic re-emission mode of mass transfer \citep{bv91}.
In this model matter flows over from the donor star ($M_2$) to
the accreting neutron star ($M_{\rm NS}$) in a conservative manner and thereafter a certain fraction, $\beta$
of this matter is ejected from the vicinity of the neutron star with the specific orbital angular momentum of
the neutron star \citep[for example, in a jet as observed in SS433, see also][]{kb99}.
Integrating the orbital angular momentum balance equation one can find the change in orbital separation during the
isotropic re-emission RLO
\citep[e.g.][]{tau96,kskd01}:
\begin{equation}
  \frac{a}{a_0}= \left( \frac{q_0(1-\beta)+1}{q(1-\beta)+1} \right) ^{\frac{3\beta-5}{1-\beta}}
                 \left( \frac{q_0+1}{q+1} \right) \left( \frac{q_0}{q} \right) ^2
\label{aa0_ire}
\end{equation}
where it is assumed that $\beta$ remains constant during the mass-transfer phase. Indeed \citet{tvs00} showed in their
Fig.~1 that in
intermediate-mass {X}-ray binary (IMXB) systems very little mass is accreted onto the neutron star since the timescale
for the mass-transfer phase is very short ($\sim\!1~{\rm Myr}$) leading to highly super-Eddington mass-transfer rates
by 3--4 orders of magnitude and hence $\beta > 0.999$. It is therefore interesting
to consider Eq.~(\ref{aa0_ire}) in the limit where $\beta \rightarrow 1$ and we find \citep[see also][]{kskd01}
for the change in orbital period:
\begin{equation}
\lim_{\beta \to 1} \left( \frac{P}{P_0} \right) =
         \left( \frac{k\,q+1}{q+1} \right) ^2 \,k^3 \, e^{3q(1-k)}
\label{PP0_limes_k}
\end{equation}
under the above mentioned assumptions and by applying Kepler's third law.
\begin{figure}
\begin{center}
  \includegraphics[width=0.35\textwidth, angle=-90]{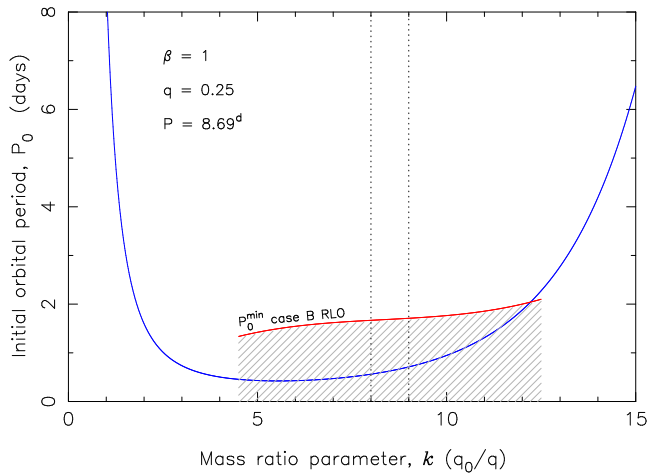}
  \caption[]{
    The blue line represents the initial orbital period of the progenitor {X}-ray binary as
    a function of the mass ratio parameter $k$ using the isotropic re-emission model for early Case~B RLO. The two vertical dotted lines indicate the
    interval of possible values of $k$ for \psr. The hatched region is excluded for Case~B RLO
    since the donor star would have filled its Roche-lobe before reaching the stage of shell hydrogen burning. The original donor star mass is
    $M_2 = k \times 0.50\,M_{\odot}$. Note, the $k$-values for Case~B RLO are larger than the $k$-values expected for
    Case~C RLO in Fig.~\ref{fig:caseC}. The reason for this is that the core of a Case~B donor star has not had time to evolve to
    the large core masses found in Case~C RLO (see Fig.~\ref{fig:coremass}),
    resulting in higher required donor masses, $M_2$ for Case~B compared to a Case~C scenario.
    }
\label{fig:caseB}
\end{center}
\end{figure}

In Fig.~\ref{fig:caseB} we demonstrate that early Case~B mass transfer is not possible to have occurred in the progenitor
binary of \psr. The constraints on $k$ for Case~B mass transfer can be found
from Fig.\ref{fig:coremass}: For Case~B mass transfer (between evolutionary epochs TAMS and the RGB)
a progenitor star of $4.0-4.5\,M_{\odot}$ is needed to yield
a core mass of $0.50\,M_{\odot}$ (the observed mass of the white dwarf in \psr). Hence, we
find $8\le k \le 9$ for this scenario. Recalling $q=M_{\rm WD}/M_{\rm NS}=0.25$ and
given the orbital period of $P=8.69$~days then, according to Eq.~(\ref{PP0_limes_k}), this would require an initial orbital
period, $P_0 \simeq 0.7\,\rm{days}$ which is not possible for Case~B mass transfer -- the minimum initial period for (early) Case~B RLO
is shown as the red line in Fig.~\ref{fig:caseB}.
In fact with such a short initial orbital period the donor star would fill its Roche-lobe radius
instantly on the ZAMS. Even if we expand the interval of donor star masses to the entire range $2.5 < M_2/M_{\odot} < 6.0$
early Case~B RLO would still not be possible to explain the parameters of \psr.
We can therefore safely rule out all variations of Case~B mass transfer (early, late and Case~BB).

\subsection{Case~A RLO -- mass transfer from a main sequence star}\label{subsec:CaseA}
In order to reproduce \psr\ via Case~A RLO 
we notice from Fig.\ref{fig:coremass} that we must at first glance require an initial donor mass of almost $M_2\simeq 5\,M_{\odot}$ in order to
end with a final white dwarf mass of $0.50\,M_{\odot}$. 
However, the evolution of Case~A RLO is somewhat complex and not straight forward to analyse analytically \citep[see][for further details]{tl11}.
The estimated TAMS core masses from
Fig.\ref{fig:coremass} are not necessarily good indicators for the final mass of the white dwarf remnants evolving from Case~A donors in {X}-ray binaries for two reasons:
1) forced mass loss from the Roche-lobe filling donor star
results in a lower core mass as the donor now evolves less massive, and 2) the formation of an outgoing hydrogen shell source
during the final phase (phase AB, see below) of the mass transfer causes the core mass to grow with the helium ashes left behind.
Therefore, to obtain the final mass of the white dwarf requires detailed numerical stellar models.
The overall effect is that 
the core mass will have grown somewhat by the time the system detaches from the RLO. Hence, the white dwarf remnant left behind is
expected to be slightly more massive than the donor core mass at the TAMS.
For this reason the ZAMS mass interval found from Fig.~\ref{fig:coremass} for a Case~A donor star of \psr\
should be considered as an upper limit
and in the following we explore donor masses down to $4.0\,M_{\odot}$. (Below this limit the WD remnant becomes too light, see Section~\ref{subsubsec:parameterspace}). 
Stars more massive than $5.0\,M_{\odot}$ could leave behind a core mass even less than $0.50\,M_{\odot}$ if the mass transfer
is initiated well before reaching the TAMS. However, these binaries would not be dynamically stable with a neutron star accretor, see Section~\ref{subsubsec:parameterspace}.

\begin{figure}
\begin{center}
  \includegraphics[width=0.35\textwidth, angle=-90]{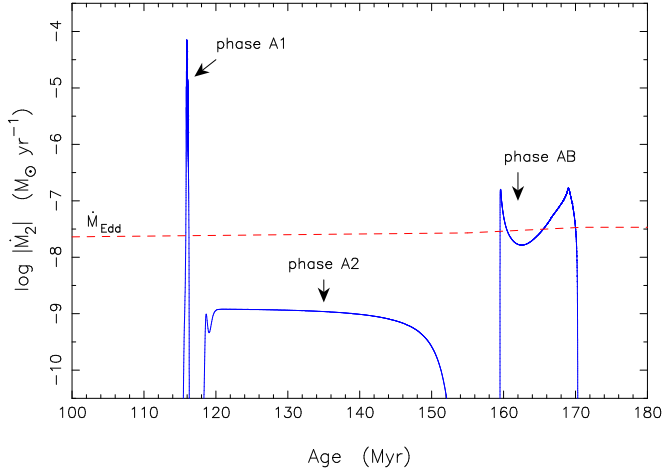}
  \caption[]{
    Case~A RLO in a binary initially made of a $4.50\,M_{\odot}$ donor star with a $1.68\,M_{\odot}$ neutron star and an initial
    orbital period of $P_{\rm orb}=2.20~{\rm days}$. The graph shows the mass-transfer rate from the donor star
    as a function of its age. Three phases of mass transfer (A1, A2 and AB -- see text) are identified.
    }
\label{fig:Mdot}
\end{center}
\end{figure}

\begin{figure}
\begin{center}
  \includegraphics[width=0.35\textwidth, angle=-90]{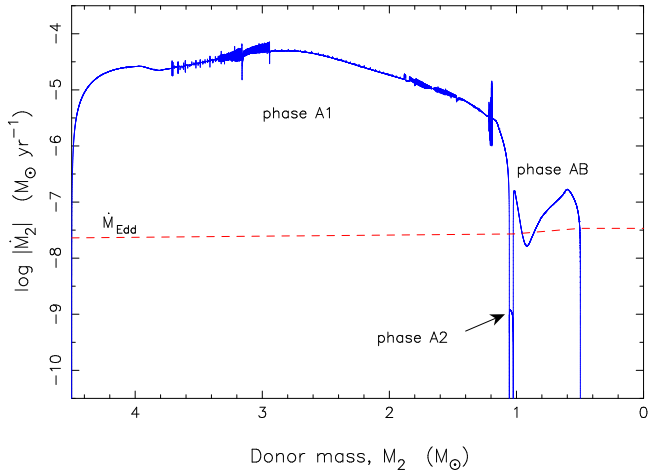}
  \caption[]{
    The mass-transfer rate as a function of the decreasing donor star mass for the stellar evolution calculation shown in Fig.~\ref{fig:Mdot}.
    Very little mass ($\sim\!0.01\,M_{\odot}$) is accreted by the neutron star during phase~A1 which proceeds on a thermal timescale. 
    }
\label{fig:MdotM2}
\end{center}
\end{figure}

Our analysis reveals the parameter space of Case~A binaries which produce the characteristic parameters
of \psr. 
Figs.~4--8 show an example of a calculation of a possible progenitor {X}-ray binary. 
This IMXB started out with an initial
donor star of mass $4.50\,M_{\odot}$ and a $1.68\,M_{\odot}$ neutron star accretor having
an initial orbital period of 2.20~days. We assumed here a convective core-overshooting of $\delta _{\rm OV}=0.20$.
In Fig.~\ref{fig:Mdot} and Fig.~\ref{fig:MdotM2} we demonstrate that the system experiences
three phases of mass transfer (hereafter denoted phases A1, A2 and AB, respectively).
In phase~A1 the mass transfer proceeds on the thermal timescale \citep[see][for further discussions on thermally unstable mass transfer]{ldwh00}.
The reason for this is the initially large mass ratio between the heavier donor star and the lighter
neutron star which causes the orbit to shrink in response to mass transfer. As mentioned earlier,
the outcome in this situation is that the donor star overfills its Roche-lobe even more -- leading to further mass loss --
and within 1~Myr it looses more than $3\,M_{\odot}$ at a rate exceeding $10^{-5}\,M_{\odot}\,\rm{yr}^{-1}$.
This rate is still less than the estimated limit for which photons are trapped, leading possibly to
rapid neutrino cooling and hypercritical accretion. It has been demonstrated, for example by \citet{fbh96} and \citet{kb99},
that it takes an accretion rate of at least a few times $10^{-4}\,M_{\odot}\,{\rm yr}^{-1}$ before the
outward diffusion of photons cannot overcome the inward advection of photons in the accreted matter.
(The exact limit is uncertain and depends, for example, on the amount and the geometry of outflows).
Although the donor star is driven out of thermal equilibrium
during phase~A1 it manages to retain hydrostatic equilibrium and the system can in this case avoid a
so-called delayed dynamical instability \citep{hw87,kw96}
which would have resulted in a common envelope and most
likely a merger event.

The final mass of the neutron star in our example is $1.99\,M_{\odot}$. The neutron star has thus accreted
a total of $0.31\,M_{\odot}$. The amount accreted in each phase is found from Fig.~\ref{fig:Mdot} by integrating
the area under the blue line which falls below the Eddington accretion limit (red dashed line).
Hardly any accretion takes place during the
very short (thermal timescale), ultra super-Eddington phase A1.
Phases A2 and AB proceed on nuclear timescales dictated by core burning of the remaining hydrogen and, later on, hydrogen shell burning, respectively.

\begin{figure}
\begin{center}
  \includegraphics[width=0.35\textwidth, angle=-90]{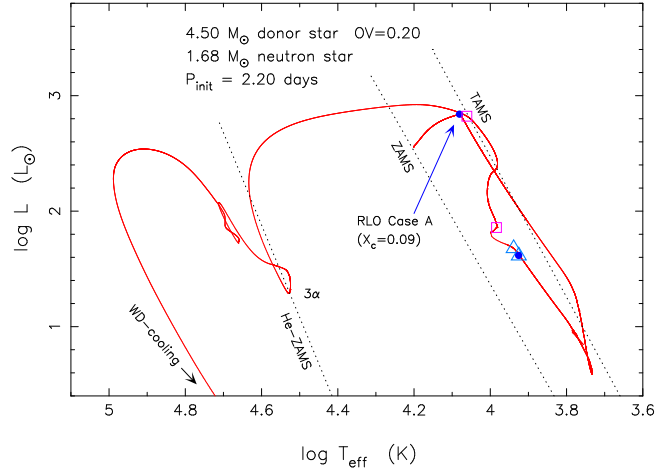}
  \caption[]{Evolution of the mass loosing donor star in the HR-diagram. 
             Starting and termination points of the three phases of mass transfer
             are shown by filled circles, open triangles and open squares, corresponding to
             phases A1, A2 and AB, respectively (cf. Fig.~\ref{fig:Mdot}). The core helium burning
             is ignited after detachment from the Roche-lobe, see text.
    }
\label{fig:HR}
\end{center}
\end{figure}

Fig.~\ref{fig:HR} shows the track of this IMXB donor star in the HR-diagram on its path to forming a
carbon-oxygen white dwarf (CO~WD) orbiting a millisecond pulsar. The Case~A RLO mass transfer is initiated at an orbital period of 2.20~days.
At this stage the core of the donor star still has a central hydrogen mass abundance of $X_c=0.09$.
The error of putting the donor star on the ZAMS is small given that the progenitor star of the neutron star (i.e. the primary star of the ZAMS binary)
most likely had a mass of at least
$20\,M_{\odot}$ (see arguments in Sections~\ref{sec:progenitor} and \ref{sec:discussion}) and hence a lifetime of less than 10~Myr, which is short compared to
the main sequence lifetime of a $4.50\,M_{\odot}$ star.
Binaries with shorter initial periods will have less evolved donor stars when entering the mass-exchange phase. This leads 
to lower helium cores masses which are then often below the threshold for igniting the triple-$\alpha$ process. Hence, these
systems will leave behind pulsars with a low-mass helium WD companion,
as first pointed out by \citet{prp02} -- for further discussion on these systems, see \citet{tl11}. 

\begin{figure}
\begin{center}
  \includegraphics[width=0.32\textwidth, angle=-90]{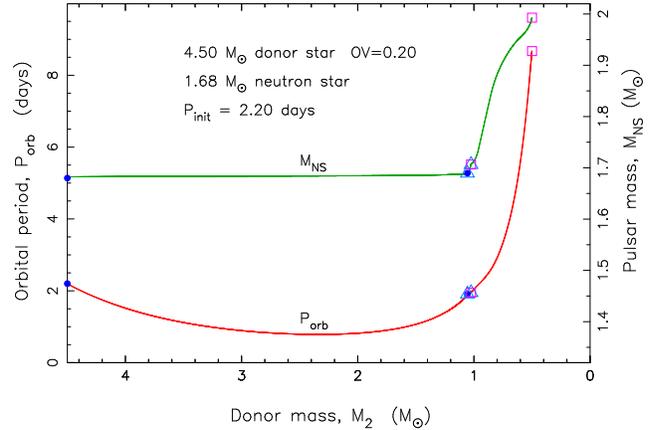}
  \caption[]{Orbital period (red line) and pulsar mass (green line) as a function of decreasing donor star mass. The symbols
             are equivalent to those defined in Fig.~\ref{fig:HR}. 
    }
\label{fig:Porb}
\end{center}
\end{figure}
When the donor star settles on the He-ZAMS its luminosity is entirely generated by the triple-$\alpha$ process in the core -- marked in Fig.~\ref{fig:HR} with the
symbol "3$\alpha$". At this stage the donor star has an age of $180\,{\rm Myr}$. The ignition of the triple-$\alpha$ process actually occurs gradually
already from ($\log {T_{\rm eff}},\,\log (L/L_{\odot}))=(4.2,\,2.9)$ when $T_{\rm core} > 10^8\,{\rm K}$, shortly after the detachment from the Roche-lobe,
and overlaps with the ceasing stages of the shell hydrogen burning.
The curly loop at $\log (L/L_{\odot})\simeq 2$ indicates the beginning of the shell helium burning phase at an age 
$t=300\,{\rm Myr}$. 

The orbital evolution is shown in Fig.~\ref{fig:Porb} where the orbital period is plotted as a function of decreasing donor star mass.
The final orbital period of our system is 8.67~days. The widening of
the orbit is quite significant in phase~AB where the mass ratio, $q$ is small.
It is also during phase~AB that the neutron star (NS) gains the majority of its accreted mass -- see green line in Fig.~\ref{fig:Porb}. 
The donor star of the {X}-ray binary Cyg~{X}-2
is an example of a hydrogen shell burning star near the end of phase~AB \citep{pr00}.
Interestingly enough, a massive neutron star ($\sim 1.8\,M_{\odot}$) has been inferred for this source as well,
see for example \citet{cck98}.
  
\begin{figure}
\begin{center}
  \includegraphics[width=0.35\textwidth, angle=-90]{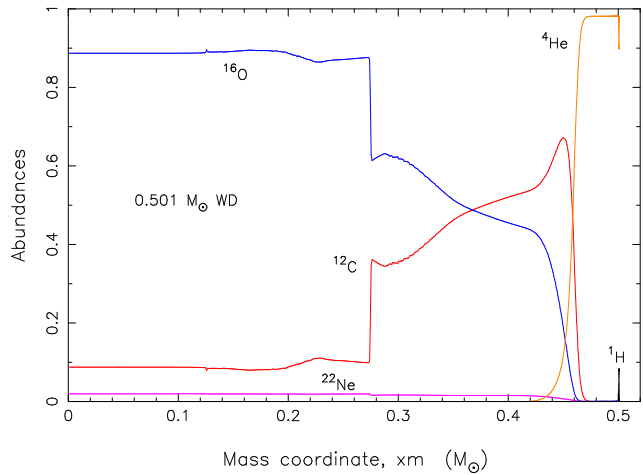}
  \caption[]{The chemical abundance structure of the CO~WD formed in the Case~A RLO shown in Figs.~\ref{fig:Mdot}-\ref{fig:Porb}.
             This profile could well resemble the
             structure of the CO~WD companion to \psr\ -- possibly applicable to a WD cooling model.
             The radius of the WD, once it settles on the cooling track, is $9500\,{\rm km}$ yielding a
             surface gravity of $\log(g)=7.9$.
    }
\label{fig:WD}
\end{center}
\end{figure}

The chemical abundance profile of the resulting CO~WD is shown in Fig.~\ref{fig:WD}.
We notice that the inner core ($\sim\!0.28\,M_{\odot}$) contains almost 90\% oxygen (mass fraction).
The CO~WD is seen to have a $\sim\!0.04\,M_{\odot}$ helium envelope and a tiny content of up to 8\% hydrogen 
in the outermost $10^{-4}\,M_{\odot}$ (amounting to a total of $1.7\times 10^{-5}\,M_{\odot}$).
The oxygen content of the WD is larger than found by \citet{lrp+11} using the {\rm MESA} code (see their Fig.~6, panel~d).
To further test the chemical composition of our WD we compared a run of a similar binary system 
($4.50\,M_{\odot}$ donor star, $1.68\,M_{\odot}$ NS, $P_{\rm orb}=2.20\,{\rm days}$, $\delta_{\rm OV}=0.20$) using
the Eggleton stellar evolution code. This computation yielded an oxygen content 
in agreement with our result (Evert Glebbeek, private communication). The reason for the different oxygen yield 
in \citet{lrp+11} could very well be caused by the use of different $^{12}{\rm C}(\alpha,\gamma)^{16}{\rm O}$ reaction rates.

To summarize, the final outcome of our example shown for Case~A evolution is a $0.501\,M_{\odot}$ CO~WD orbiting a
$1.99\,M_{\odot}$ (millisecond) pulsar with
an orbital period of $8.67$~days -- almost exactly in agreement with the observed parameters of \psr, see Table~\ref{table:param}.

\subsubsection{Permitted parameter space for Case~A RLO leading to the formation of \psr}\label{subsubsec:parameterspace}
\begin{figure}
\begin{center}
  \includegraphics[width=0.35\textwidth, angle=-90]{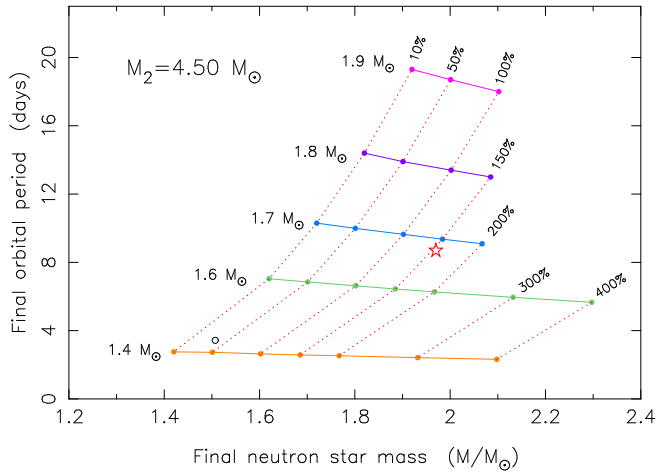}
  \caption[]{
    The final orbital period as a function of final neutron star mass for a grid of {X}-ray binaries evolving
    from a $4.50\,M_{\odot}$ donor star through Case~A RLO. In all models the CO~WD is formed with a mass of about $0.51\pm0.01\,M_{\odot}$.
    The initial orbital period was in all cases about 2.2~days, corresponding to a core hydrogen content of $\sim\!10$\% at the time of RLO.
    The variables are the {\it initial} neutron star mass (solid lines) and the accretion
    efficiency (dotted lines).
    The observed values of \psr\ are shown with a red star.
    Our calculations show that indeed \psr\ could have evolved from a $4.5\,M_{\odot}$ donor star
    and a neutron star born with a mass of $\sim\!1.7\,M_{\odot}$, accreting at an efficiency of 150\% -- see text.
    }
\label{fig:grid4.50}
\end{center}
\end{figure}

We have demonstrated above that both Case~C and Case~A RLO during the {X}-ray phase can reproduce the observed parameters of \psr\
for suitable initial masses of the two components and their orbital period. 
In order to search the entire parameter space of Case~A systems we explored a range of binaries by altering the stellar masses,
the orbital period and the accretion efficiency.
In Fig.~\ref{fig:grid4.50} we show the grid of resulting NS+WD systems in the final orbital period versus final neutron star mass plane.
This plot was obtained by varying the initial mass of the neutron star as well as the accretion efficiency for a fixed value of the
donor star mass, $M_2=4.50\,M_{\odot}$ at the onset of the {X}-ray phase. 
It is obvious that the final mass of the neutron star is a growing function of its initial mass as well as the efficiency of accretion.

In order to be able to compare with previous work we have in this plot defined the accretion efficiency as a value in percent of the
Eddington mass accretion limit ($\dot{M}_{\rm Edd}$) for pure hydrogen on a neutron star with a radius of 10~km, 
such that a value of 100\% corresponds to the canonical accretion rate of
$1.5\times10^{-8}\,M_{\odot}\,{\rm yr}^{-1}$.  A value larger than 100\% corresponds to either accretion at slight super-Eddington rates or
accretion of matter with a larger mean molecular weight per electron (e.g. an accretion efficiency value of 200\% corresponds to accreting
pure helium at the Eddington limit).\\
Many of the grid points in Fig.~\ref{fig:grid4.50} are not obtained from actual stellar evolution calculations. For example, the evolution
leading to grid points based on an
initial neutron star mass of $1.4\,M_{\odot}$ were dynamically unstable in our models, leading to runaway mass transfer \citep[see below, and also][]{prp02}.
Nevertheless, one can still compare with the calculations in \citet{prp02}. Using a neutron star with an initial
mass of $1.4\,M_{\odot}$ orbiting a donor star of mass $4.5\,M_{\odot}$ with an initial orbital period of 2.38~days, and assuming an accretion efficiency of 
50\%, these authors end up with a NS+WD
binary with a final neutron star mass of $1.507\,M_{\odot}$, a white dwarf mass of $0.471\,M_{\odot}$ and an orbital period of 3.43~days. (Also in their work
they find that such an {X}-ray binary is barely on the edge of stability and note that this system may be dynamically unstable). The result
of their calculation is shown in our figure with an open black circle and the agreement with our result 
is indeed quite good (cf. the orange neighbour point in our grid just below their point).
Our result is based on one
of our calculations with a $4.50\,M_{\odot}$ donor star and a neutron star mass high enough to avoid a dynamical instability, for example of mass $1.7\,M_{\odot}$.
The effect of changing the neutron star mass and/or the accretion efficiency can easily be found analytically from an extrapolation of
our calculated model using Eq.~(\ref{aa0_ire}) for each of the three phases of mass transfer by adapting the new values of
$\beta$, $q$ and $q_0$. The underlying assumption that the
amount of mass transfered from the donor star remains roughly constant (i.e. independent on the neutron star mass) has been tested by us and shown to be correct. This was done by
directly comparing the result of an extrapolated model with a calculated model. See \citet{tl11} for further discussion.\\
In our stellar evolution code the Eddington accretion limit (i.e. the accretion efficiency) depends on the chemical composition of the accreted matter as well as the 
radius of the neutron star, both of which are time dependent -- see description in Paper~II, and the red dashed lines
in Figs.~4--5. 

It is important to notice from Fig.~\ref{fig:grid4.50} how the final orbital period is correlated with the initial mass of the neutron star
(increasing in value upwards in the grid diagram, see solid lines).
The grid clearly shows that, in the frame of Case~A RLO, the neutron star in \psr\ cannot have been born with the canonical birth mass of about $1.3\,M_{\odot}$. 
This conclusion was also found by \citet{lrp+11}.

Using our stellar evolution code we find that our IMXBs are only stable against dynamical mass transfer for initial mass ratios
up to $q_0\simeq 2.7-3$, e.g. corresponding to initial donor masses at most $3.5-4.0\,M_{\odot}$ for a $1.3\,M_{\odot}$ neutron star and
(what is important for \psr) donor stars up to $5.0\,M_{\odot}$ for a $1.7\,M_{\odot}$ neutron star. Therefore we adapt $5.0\,M_{\odot}$
as the upper limit for the initial mass of the donor star. The lower limit for the mass of the donor star is constrained by the mass of
the CO~WD in \psr. We find a lower limit of about $4.0\,M_{\odot}$ (a $3.5\,M_{\odot}$ donor star in the region of relevant initial orbital periods 
leaves behind a WD mass of only
$0.39\,M_{\odot}$ which is 20\% smaller than needed for \psr).\\
The effect on the final orbital period and neutron star mass, imposed by changing only the donor star mass and keeping all other parameters fixed, 
can be visualized by moving the entire grid in Fig.~\ref{fig:grid4.50} up or down 
for a less massive and a more massive donor star, respectively.
We find that a $4.0\,M_{\odot}$ donor star would need to be in a binary with a neutron star of initial mass of $1.55\,M_{\odot}$ in order to reproduce \psr.
The $5.0\,M_{\odot}$ donor star would need a neutron star of initial mass of $1.77\,M_{\odot}$ in order to reproduce \psr. In both cases the
required accretion efficiency value is about 160-170\%.
However, the precise limits depend on, for example, the uncertain strength and the assumed underlying physics of the tidal torques and 
resulting spin-orbit couplings \citep[which may help to stabilize the orbital evolution in {X}-ray binaries with even higher mass ratios,][]{ts01}.

There is evidence from comparison of numerical binary stellar evolution calculations and observations of recycled pulsars
that the accretion efficiency in some cases is rather low \citep[see Fig.~4b, \S5.7 and \S6.4 in][]{ts99}, possibly as a result
of the propeller effect and/or accretion disk instabilities. A low accretion efficiency would have the implication
for \psr\ that it was born with an even higher mass. For a $4.5\,M_{\odot}$ donor star and assuming an accretion efficiency
of 50\% (according to the definition given earlier) this would result in a neutron star birth mass of about 
$1.9\,M_{\odot}$. However, in this case one can only reproduce the observed orbital period of 8.69~days by increasing
the specific orbital angular momentum of the lost material beyond the value expected for matter in the vicinity of the accreting neutron star
(i.e. by matter ejected from a location elsewhere within the binary system). To illustrate this 
one can see from Fig.~\ref{fig:grid4.50} that an initial neutron star mass of $1.9\,M_{\odot}$ and an accretion
efficiency of 50\% would yield a final orbital period of about 18~days (much larger than observed in \psr). This problem can be solved if one instead assumes 
that the material lost from the system has the {\it average} specific orbital angular momentum of the binary.
The reason is that the orbit widens mainly during the mass-transfer phase AB, when the accreting neutron star is more massive than the donor star,
and in our model the ejected material has the low specific orbital angular momentum of the neutron star during this phase.

\section{Evolution of progenitor binaries from the ZAMS to the {X}-ray phase}\label{sec:progenitor}
\begin{figure}
\begin{center}
  \includegraphics[width=0.45\textwidth, angle=0]{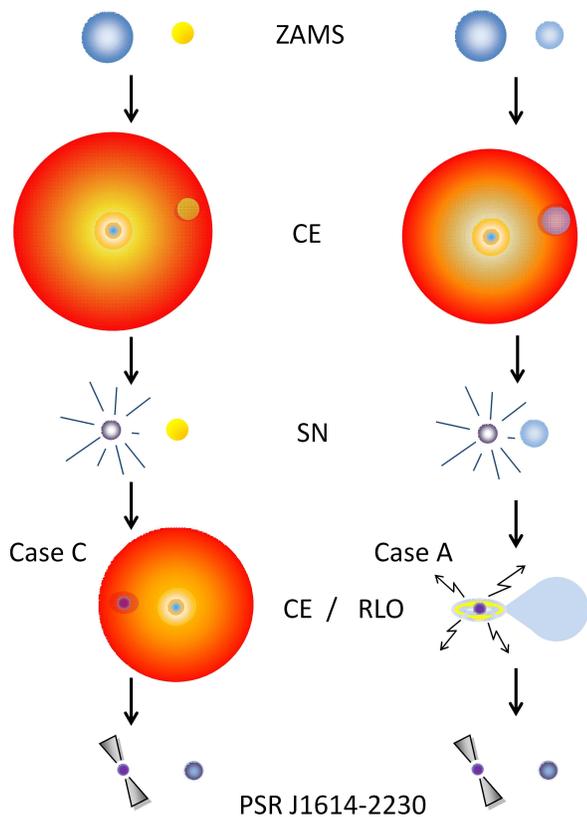}
  \caption[]{An illustration of the progenitor evolution leading to the formation
             of \psr\ for both Case~A and Case~C. Only a few
             evolutionary epochs are shown for simplicity.
    }
\label{fig:vdhcartoon}
\end{center}
\end{figure}

In the previous section we presented evidence for two different formation scenarios (hereafter simply called Case~A and Case~C) for the
formation of \psr. Hence, we know the required parameters at the onset of the RLO, for the {X}-ray binary containing a neutron star 
and a non-degenerate star, and one can then try to calculate backwards and estimate the initial configuration of ZAMS binaries which may eventually 
form a system like \psr. 
Our brief description presented here is only qualitative. An analysis including, for example, dynamical effects of asymmetric supernovae (SNe) 
is rather cumbersome. To obtain a set of more detailed and finetuned parameters of
the progenitor binaries one would need to perform a population synthesis (which is beyond the scope of this paper).
Nevertheless, below we present the main ideas. The results of our simple analysis are shown in Table~\ref{tab:progenitor} and illustrated
in Fig.~\ref{fig:vdhcartoon}. 

A mass reversal between the two interacting stars during the evolution from the ZAMS to the {X}-ray phase can be ruled out
in both scenarios
as a consequence of the large difference in ZAMS mass between the two stars. The secondary star, the donor star of the {X}-ray binary,
had a mass of $2.2-2.6\,M_{\odot}$ or $4.50\,M_{\odot}$, respectively. The primary star had at least a mass above the threshold of
$\sim\!10\,M_{\odot}$ for producing a NS in a close binary.
In Section~\ref{subsec:NSmass_theory} we argue that the mass of the progenitor of the NS in \psr\ was most likely more than
$20\,M_{\odot}$ since it left behind a massive NS.
Hence, there is no doubt that the NS is the remnant of the
original primary star, $M_1$ (i.e. the initially more massive) in the ZAMS binary and the donor star in the {X}-ray phase thus descends from
the secondary ZAMS star, $M_2$. 
Based on this argument, due to the small mass ratio $M_2/M_1 \simeq 0.1-0.25$ we would expect the 
progenitor binary to have evolved through a CE and spiral-in phase in both Case~A and Case~C on their path
from the ZAMS to the SN stage. The initial ZAMS orbital period was then probably quite large ($>10^3~{\rm days}$)
in order to let the primary star evolve to a late evolutionary stage before the onset of the CE.
This would both facilitate the ejection of the envelope -- which is more loosely bound at late evolutionary stages --
and ensure that the helium core of the neutron star progenitor evolved "clothed" -- see Section~\ref{subsec:NSmass_theory}. 
We will now briefly discuss each of the two scenarios: Case~C and Case~A. 
\begin{table*}
\center
\caption{Evolution and characteristics of the possible progenitor binaries of \psr. 
         The different columns correspond to the different cases of RLO
         in the {X}-ray binary phase. In this table the evolution with time goes
         from top to bottom (see also Fig.~\ref{fig:vdhcartoon}).}
\begin{tabular}{lccccl}
\hline
\noalign{\smallskip}
    {Initial ZAMS}                             & $\downarrow$ & $\downarrow$ & $\downarrow$ & $\downarrow$          & {Comments} \\
\hline
\hspace{0.3cm} Primary mass ($M_{\odot}$)      &  20--25      & --           & 20--25       & 20--25                 &  In all cases the evolution from the \\
\hspace{0.3cm} Secondary mass ($M_{\odot}$)    &  2.2--2.6    & --           & 4.0--5.0     & 4.50                   &  ZAMS to the X-ray phase goes \\
\hspace{0.3cm} Orbital period (days)           & $10^3$       & --           & $10^3$       & $10^3$                 &  through a (first) CE-phase \\
\hline
\noalign{\smallskip}
    {Initial X-ray binary}                     & {\bf Case C} & {\bf Case B} & {\bf Case A} & {\bf Case A$^*$}       & $^*$Our example shown in this paper \\
\hline
\hspace{0.3cm} Neutron star mass ($M_{\odot}$) &  1.97        & --           & 1.55--1.77   & 1.68                   &  The results of this paper \\
\hspace{0.3cm} Donor mass ($M_{\odot}$)        &  2.2--2.6    & --           & 4.0--5.0     & 4.50                   &  \\
\hspace{0.3cm} Orbital period (days)           &  $>10^{2\,**}$  & --        & 2.0--2.3     & 2.20                   &  $^{**}$ Depending on details of the first CE\\
\hline
\noalign{\smallskip}
    {Binary millisecond pulsar}                & $\downarrow$ & $\downarrow$ & $\downarrow$ & $\downarrow$           & {\bf \psr}  \\
\hline
\hspace{0.3cm} Pulsar mass ($M_{\odot}$)       &  1.97        & --           & 1.95--2.05   & 1.99                   & 1.97  \\
\hspace{0.3cm} White dwarf mass ($M_{\odot}$)  &  0.50        & --           & 0.47--0.53   & 0.501                  & 0.500 \\
\hspace{0.3cm} Orbital period (days)           &  0.1--20     & --           & 3--16        & 8.67                   & 8.69  \\
\hline
\end{tabular}
\label{tab:progenitor}
\end{table*}

\subsection{Case~C progenitor binary}\label{subsec:CaseCprogenitor}
The aim here is to obtain a $2.2-2.6\,M_{\odot}$ non-degenerate AGB star orbiting a NS (see Section~\ref{subsec:CaseC}).
Hence, the binary must have been very wide ($\sim\!10^3$~days) following the SN in order to allow the donor star
to ascend the AGB before initiating mass transfer. A wide post-SN orbit could have been the result of a large
kick imparted to the newborn NS at birth \citep{ll94}. However, tidal interactions in a highly eccentric orbit
may significantly reduce the orbital separation \citep{sut74} thus preventing the companion star from ascending the AGB before it fills its Roche-lobe.
The other alternative, namely that the wide
post-SN orbit simply reflects that the pre-SN orbit was wide too is also possible, but not very likely either. Such a wide system would rarely
survive any kick imparted to the newborn neutron star -- and a small kick originating from an electron capture supernova
would be in contradiction with the high neutron star mass \citep{plp+04,vdh04}. Furthermore, the expected outcome 
of the first CE~phase (between the ZAMS and SN stages) is not a wide system.
The reason is that the binding energy of the massive primary star's envelope is too large, on an absolute scale,
to allow for an early ejection.
Hence, there could probably not have been a "mild in-spiral" as a result of
an easy ejection of the envelope resulting in only a modest conversion of orbital energy and allowing the
orbit to remain fairly wide. An alternative possibility is that the initial orbit was so
wide that the two stars did not exchange mass during the giant phase of the primary star. The subsequent
kick in the SN then shot the newborn NS into a closer orbit around the secondary star \citep{kal98}. However, this scenario
would require a very fortunate finetuning of both the kick magnitude and the direction, making it unlikely too.

\subsection{Case~A progenitor binary}\label{subsec:CaseAprogenitor}    
To produce an {X}-ray binary with an orbital period of only 2.20~days and a $4.50\,M_{\odot}$ donor star
seems more likely compared to the case described above. The short orbital period, both before
and after the SN, is a simple consequence of the in-spiral during the first CE-phase when the
primary star was a giant. 
We therefore conclude at this stage, that based on binary evolution considerations
\psr\ seems more likely to have evolved from a Case~A RLO {X}-ray binary and that the initial ZAMS system
was composed of a $\ge20\,M_{\odot}$ primary with a $4-5\,M_{\odot}$ secondary in a wide orbit. 
In the next section we discuss the stellar evolution of the NS progenitor in much more detail and
in Paper~II we discuss progenitor systems based on the spin-up of the neutron star.

\section{Discussion}\label{sec:discussion}
The precise measurement of the high neutron star mass in \psr\ 
leads to interesting
implications for the nuclear physics behind the equation-of-state \citep[e.g.][]{lp11}.
Equally important, the result has caused renewed interest
in modelling close binary evolution -- in particular the mass-transfer phase. 
In a recent paper \citet{kkt11} discuss the possibility of "{\it alternative} evolution" in order to
explain the observed mass of \psr. 
However, in Section~\ref{sec:masstransfer} of this paper
we have demonstrated
that this is not required and \psr\ may have followed standard evolution paths expected from stellar astrophysics.\\
The challenge in reproducing massive binary millisecond pulsars with a {CO} white dwarf companion (like \psr)
is to get all three fundamental observable parameters correct: the masses of the two compact objects
and the orbital period.
In this paper we have demonstrated a methodical approach to do this involving all three mass-transfer scenarios
(RLO Cases A,~B and C).

In a recent paper \citet{lrp+11} systematically computed the evolution of a large number of 
Case~A and early Case~B IMXB binaries and applied their results
to understand the formation of \psr. They conclude that a system like \psr\ requires a minimum initial neutron star
mass of at least $1.6\pm0.1\,M_{\odot}$, as well as an initial donor mass of $4.25\pm0.10\,M_{\odot}$ and
orbital period of $\sim\!49\pm2\,\rm{hr}$ ($2.05\pm0.1\,{\rm days}$). In general their Case~A results are in fine agreement
with our Case~A results. The main difference is their rather narrow range of required donor star masses ($4.25\pm0.10\,M_{\odot}$)
compared to our wider interval of 4.0--5.0$\,M_{\odot}$. This minor discrepancy could arise from using different values of the 
convective core-overshooting parameter, the mixing length parameter and/or the chemical composition of the donor star\footnote{These parameters 
are not stated in their present publication.}. 
However, it is interesting to notice the broad agreement in the final results given that the stellar evolution codes are different.

It is important to emphasize that other observed recycled pulsars with a {CO}~WD companion and $P_{\rm orb} \simeq 5-15$~days
(such as PSR~J0621+1002 and PSR~J2145$-$0750) are, in general, not expected to have a massive neutron star.
\citet{tvs00} demonstrated that these systems can be reproduced by early Case~B RLO with a typical $1.3\,M_{\odot}$
neutron star and \citet{vdh94b} argued for a CE evolution scenario using a neutron star mass of $1.4\,M_{\odot}$. 
Systems with a massive neutron star, like \psr\ investigated here, require the neutron star to be born massive
-- possibly followed by an extended phase of mass transfer allowing for significant further accretion of matter.

\subsection{Neutron star birth masses in pulsar binaries}\label{subsec:NSmass}
The interval of known radio pulsar masses ranges from $1.17\,M{_\odot}$ in
the double neutron star binary PSR~J1518+4909 \citep[3-$\sigma$ upper limit,][]{jsk+08}
to $1.97\,M_{\odot}$ in \psr, discussed in this paper.
The most massive of the non-recycled companions in double neutron star systems
is the unseen companion in PSR~1913+16 which has a mass of $1.389\,M_{\odot}$ \citep{wnt10}.
Interestingly, the observed pulsar in this binary is the most massive of the (mildly) recycled pulsars
detected in any of the ten double neutron star systems. It has a mass of $1.440\,M_{\odot}$. However, the relatively slow
spin period of this pulsar (59~ms) hints that only about $10^{-3}\,M_{\odot}$ was needed in the recycling process 
(see Paper~II) and thus $1.44\,M_{\odot}$ is the previously known upper limit
derived for the {\it birth} mass of any neutron star detected in a binary pulsar system. 
Only a few of the $\sim\!120$ binary pulsars with WD companions have measured masses -- see Paper~II -- and just a handful 
of these are more massive than $1.44\,M_{\odot}$. 
But even in those cases the mass determinations are often very inaccurate and also include the mass accreted from 
the progenitor of their WD companion. 
However, in this paper we have demonstrated that the birth mass of the neutron star in \psr\ is at least $1.7\pm0.15\,M_{\odot}$.
This result is important for understanding the physics of core collapse supernova and neutron star formation. 
 
\subsection{Predictions from stellar evolution and SN~explosion models}\label{subsec:NSmass_theory}
The difficulty of predicting an upper limit for the birth mass of a neutron star is mainly caused by unknown details of stellar evolution 
and explosion physics, as well as the neutron star equation-of-state. 
For example, internal mixing processes, wind mass loss and key nuclear reaction rates are still not known accurately.
Additionally, even models
with the same input physics predict a rather erratic behaviour of the final iron core mass as a function
of ZAMS~mass -- cf. Fig.~17 in \citet{whw02} and Fig.~4 in \citet{tww96}. Moreover, rotation and metallicity (and perhaps magnetic fields) affect
the final remnant mass too. Rotation induced chemical homogeneous evolution will cause rapidly rotating stars to develop larger cores \citep[e.g.][]{yln06}
and a low metallicity content also causes stars to develop somewhat larger cores as a result of reduced wind mass loss \citep{vdl01} --
see also \citet{lks+10} and \citet{mdl+11} who discuss the relatively high number of black hole HMXBs at low metallicities 
(compared to HMXBs with a neutron star companion).

As for the explosion physics there are, for example, uncertainties in deriving the explosion energy 
as function of the pre-supernova structure, and 
in the determination of the mass cut separating the initial compact remnant from the ejected matter.
Also the amount of fall back material is uncertain. 

Despite the above-mentioned uncertainties we can identify
two main factors which determine the remnant mass of a given early-type star: 
1) its ZAMS~mass, and 2) whether or
not it looses its hydrogen-rich envelope  (e.g. as a result of mass transfer in a binary) 
before or during 
core helium burning. Both of these factors influence the carbon abundance at core helium exhaustion which
plays a crucial role for the subsequent carbon, neon and oxygen burning stages which again determines the size of the 
silicon and iron cores (and thus the mass of the newborn neutron star).

\subsubsection{To burn carbon convectively or not -- the role of the $^{12}{\rm C}/^{16}{\rm O}$-ratio at central helium depletion}\label{subsubsec:COratio}
Following \citet{bhl+01},
the $^{12}{\rm C}/^{16}{\rm O}$-ratio at central helium depletion
is determined by the competition between the formation of carbon via the triple-$\alpha$ process ($3\,\alpha\!\rightarrow \!^{12}\rm{C}$) 
and the destruction of carbon, mainly via $\alpha$-capture: $^{12}{\rm C}(\alpha,\gamma)^{16}{\rm O}$.
More massive stars
perform core helium burning at higher temperatures and
lower densities, which decrease the net carbon yield.
In stars initially more massive than about $19-20\,M_{\odot}$ 
the resulting core carbon abundance is too small (in mass fraction $C_c \le 0.15$)
to provide a long lasting convective core carbon burning phase \citep{ww95}. 
The short lasting (radiative) core carbon burning phase results in
less energy carried away by neutrinos. As a consequence, the entropy in the core remains fairly high
which leads to more massive pre-SN cores.
A low carbon abundance also causes the 
carbon burning shells to be located further out which increases the size of the carbon-free core and 
thus increases the mass of the iron core to form.
Single stars with ZAMS masses less than $20\,M_{\odot}$, on the other hand, 
deplete core helium burning with a relatively high 
$^{12}{\rm C}/^{16}{\rm O}$-ratio. These stars 
undergo significant convective carbon burning 
which subsequently leads to relatively small iron cores and hence low-mass neutron stars.


\subsubsection{Single stars / wide orbit binaries (Case~C RLO)}\label{subsubsec:single}
Based on the core carbon burning dichotomy discussed above, \citet{tww96} found  
a bimodial distribution of neutron star birth masses with narrow peaks at
$1.28\pm0.06\,M_{\odot}$ and $1.73\pm0.08\,M_{\odot}$ (gravitational masses),
for single star progenitors (type$\,$II~SNe) below and above the critical
ZAMS mass of $\sim\!20\,M_{\odot}$.
Based on assumptions of rather soft equations-of-state for nuclear matter at high densities
\citet{bhl+01} did not predict high-mass neutron stars ($\sim\!1.7\,M_{\odot}$) to be formed and therefore concluded that
all single ZAMS stars in the interval $20-25\,M_{\odot}$ end their lives as black holes -- see also \citet{fry99,fry06}. However, given the discovery
of the massive pulsar \psr\ this picture has to change. Following our work in this paper the birth mass of this neutron star
is at least $1.7\pm0.15\,M_{\odot}$ and given the expected wide orbit of its progenitor star (see Section~\ref{sec:progenitor})
we conclude that it formed from an original ZAMS star of mass
$20-25\,M_{\odot}$ which underwent Case~C RLO (leading to the CE-phase prior to the SN -- see Fig.~\ref{fig:vdhcartoon}). 
Hence, in general we expect $20-25\,M_{\odot}$ single (or wide orbit 
stars) to form high-mass neutron stars. 
Beyond $25\,M_{\odot}$, the large binding energies of the mantle is expected to result in large fall back and the production of a black hole. 

For the sake of completeness, at the other end of the scale where stars with an initial mass of about $10\,M_{\odot}$ end their lives 
the minimum gravitational mass expected for a newborn neutron star is
about $1.25\,M_{\odot}$ for an electron capture SN \citep[e.g.][]{plp+04}.
However, for single stars, the initial mass range for stars to undergo electron capture SN
is very small, such that only a few percent of all core collapse supernovae
are expected to go through this channel \citep{phlh08}.
It is interesting to notice that small iron core collapse SNe of type$\,$II also seem to allow for the formation
of $\sim\!1.15\,M_{\odot}$ neutron stars \citep{tww96}.

To summarize, one would expect three peaks in the distribution of initial neutron star masses from the evolution of single stars 
or stars in wide orbit binaries \citep[as pointed out by][]{vdh04}:
one (small) peak at $\sim\!1.25\,M_{\odot}$ from electron capture SNe of $8-10\,M_{\odot}$ stars (leading to a small kick), 
one peak at $1.25-1.4\,M_{\odot}$ from iron core collapse SNe  
of $10-20\,M_{\odot}$ stars and one high-mass peak at $\ge1.7\,M_{\odot}$ from iron core collapse SNe  
of $20-25\,M_{\odot}$ stars. Above a ZAMS mass of $25\,M_{\odot}$ single stars are expected to form black holes -- unless the metallicity is very high (above solar). 
Rapid rotation, on the other hand, may produce black holes from stars less massive than $25\,M_{\odot}$. 

\subsubsection{Close binary stars (Case~A and Case~B RLO)}\label{subsubsec:binary}
The progenitors of neutron stars in close binaries loose their hydrogen envelope as a result of 
mass transfer (cf. Section~\ref{sec:masstransfer}) and thus end up as type$\,$Ib/c~SNe. 
As shown by \citet{bhl+01}, see also \citet{wl99},
stars which loose their envelope before or early during core helium burning 
(as in Case~A/B RLO in a close binary) evolve as "naked" helium stars. 
There are four reasons why "naked" helium stars, at least at solar metallicity, develop small cores.
Firstly, the lack of a hydrogen burning shell on top of the helium core prevents the core
mass from growing during core helium burning. 
Secondly, the lack of this shell also prevents the convective part of the core from growing during core helium burning.
In single stars, the growing convective core brings in fresh helium, which
during late core helium burning is mostly used to convert carbon to oxygen. 
"Naked" helium stars therefore end core helium burning with a much higher carbon abundance
(compared to single or wide orbit stars which evolve helium cores as "clothed"), and
this will lead to lower mass iron cores as discussed above. 
Thirdly, helium stars emit strong winds 
as so called Wolf-Rayet stars, which reduces the final CO-core mass. 
And finally, not only are the core masses of "naked" helium stars reduced,
compared to the situation of single or wide orbit stars, 
also the masses and the binding energies of the surrounding envelope are reduced, which leads to
a corresponding reduction in the amount of material that falls back after the SN explosion \citep{ww95}.
All these effects lead to smaller neutron star masses in close binaries where the progenitor
of the neutron star evolved through Case~A/B RLO,
compared to formation from a single star or wide binary (Case~C RLO) evolution.  

\citet{tww96} calculated remnant masses from explosions of the pre-SN helium star models of \citet{wlw95}. 
They estimate these models correspond
to ZAMS masses up to $35\,M_{\odot}$ for stars in close binaries which 
loose their hydrogen envelope before or early during core helium burning.
\citet{tww96} predict an upper limit for the gravitational mass of these neutron stars 
to be about $1.4\,M_{\odot}$ (see their Fig.~6c for type$\,$Ib~SN).
Depending on the details of the fall back of material these values could 
be as high as $\le1.6\,M_{\odot}$.
In fact, according to \citet{bhl+01} even stars initially as massive as 
$40-60\,M_{\odot}$ may end their lives as a neutron star if they loose
their envelope early in their evolution (Case~A or Case~B RLO).\\
At the low end of progenitor masses, \citet{plp+04} showed that electron capture supernovae
from stars in close binary systems may originate from a much broader mass range
than in the single star (or wide binary) case due to the avoidance of the second 
dredge-up phase.

In a recent paper \citet{ywl10} investigated the evolution of stars 
from binaries leading to type$\,$Ib/c~SNe
using the most recent theoretical models of (reduced) Wolf-Rayet winds. 
The result is larger final helium core masses
which may lead to somewhat larger iron core masses and thus larger post-SN remnant masses.
An investigation of this question is currently in progress.

All in all, we see that the expected neutron star mass distribution from close binary stars 
is very different from that expected from single stars. In general, close binaries 
are thought to produce lower mass neutron stars.
In this context, our conclusion that
that the first mass transfer in the progenitor evolution of \psr\ was of Case~C (a wide binary)
is fully consistent with the high initial neutron star mass of $1.7\,M_{\odot}$
we derived for this system above. 

\subsection{Ramifications from \psr, Vela~X-1 and the Black-Widow pulsar}\label{subsec:ramifications}
Mass determinations of Vela~{X-1} \citep{bkv+01,rom+11} suggest
that this neutron star has an observed mass of $1.77\pm0.08\,M_{\odot}$. The companion star to Vela~{X-1}
is a B0.5~Ib supergiant (HD~77581) with a mass of about $23\,M_{\odot}$ which implies that
the present mass of the neutron star is very close to its birth mass. (Even a hypothetical strong wind accretion  
at the Eddington limit would not have resulted in accretion of more than about $10^{-2}\,M_{\odot}$ given
the short lifetime of its massive companion). 
We therefore conclude that not only was the neutron star in \psr\ born massive ($1.7\pm0.15\,M_{\odot}$) 
also the neutron star Vela~{X-1} was born with a mass of at least $1.7\,M_{\odot}$. 
Furthermore, both of these neutron stars were produced from progenitors with a ZAMS mass above $20\,M_{\odot}$
and these progenitors did not loose
their envelope before core helium burning exhaustion (i.e. they evolved via Case~C~RLO prior to the SN
-- not to be confused with the RLO during the later {X}-ray phase discussed in Section~\ref{sec:masstransfer}).
Unlike \psr, which had a relatively low- or intermediate-mass companion, Vela~{X-1} has a massive companion
and this system is therefore not expected to have evolved through a CE prior to the SN.

A recent analysis by \citet{vbk11} of the so-called Black-Widow pulsar yielded 
a mass of $2.4\,M_{\odot}$.
Although the uncertainties of this result are rather large, such a mass would be difficult to explain
if the neutron star was born with the canonical mass of about $1.4\,M_{\odot}$.

\subsection{Future observational constraints}\label{subsec:NSmass_future_obs}
Future observations could push the empirical upper limit of the possible neutron star birth mass to even higher values. 
This could, for example, be achieved by measurements of a binary radio pulsar which reveal a massive neutron star in a double neutron star binary or, perhaps more likely, 
in a very tight binary with an {O-Ne-Mg}~WD. In both cases the evolutionary timescale of the progenitor
of the last formed compact object would be so short that no substantial amount of mass could be accreted by the
mildly recycled pulsar. Hence, in these cases the observed mass of the pulsar would be almost equal to its birth mass.
Also detection of an even more massive neutron star than Vela~{X-1} in a HMXB system would push the limit upwards.

We note that it is much more difficult and uncertain to estimate the birth mass of a pulsar in a close binary with a low-mass
helium~WD companion. Even in binaries with $P_{\rm orb}<1$~day where the mass transfer rate is expected to have been sub-Eddington \citep[e.g.][]{ts99}
it is difficult posterior to estimate the effects on the accretion rate, and thus infer the birth mass of the neutron star, 
due to the poorly known occurrence of accretion disk instabilities and strength of the propeller effects -- cf. further discussion in Paper~II.

\section{Conclusions}\label{sec:conclusions}
We have investigated the formation of \psr\ by detailed modelling of the mass exchanging {X}-ray phase of the progenitor system.
We have introduced a new analytic parameterization for calculating the outcome of either a CE evolution or the highly super-Eddington isotropic re-emission mode,
which depends only on the present observable mass ratio, $q$ and the ratio between the initial and final donor star mass, $k$.
Using a detailed stellar evolution code we calculated the outcome of a number of IMXBs undergoing Case~A RLO.
Based on the orbital dynamics and observational constraints on the stellar masses we find that \psr\ could have evolved from a neutron star with either
a $2.2-2.6\,M_{\odot}$ asymptotic giant donor star through a CE evolution (initiated by Case~C RLO), or from a 
system with a $4.0-5.0\,M_{\odot}$ donor star via Case~A RLO.
Simple qualitative arguments on the evolution from the ZAMS to the {X}-ray phase suggest that Case~A is the most likely of the two scenarios.
The methods used in this paper, for RLO Cases A,~B and C, could serve as a recipe for investigations of the progenitor system of other massive binary millisecond pulsars
with heavy white dwarf companions to be discovered in the future.\\
We conclude that the neutron star in \psr\ was born significantly more massive ($1.7\pm0.15\,M_{\odot}$) than neutron stars
found in previously known radio pulsar binaries -- a fact which is important for understanding stellar evolution of massive stars in binaries
and the explosion physics of core collapse SNe.
Finally, based on this high value for the neutron star birth mass we argue that the progenitor star of \psr\ 
had a ZAMS mass of $20-25\,M_{\odot}$ and did not loose its envelope before core helium exhaustion.\\
In Paper~II we continue the
discussion of the formation of \psr\ in view of the spin-up process and include general aspects of accretion onto a neutron star during the
recycling process and apply our results to other observed millisecond pulsars.


\section*{Acknowledgments}
We thank Sung-Chul Yoon for help and discussions on the stellar evolution code 
and the referee Marten van~Kerkwijk for suggesting to elaborate on
the expected masses of neutron stars formed under various conditions.
\bibliographystyle{mn2e} 
\bibliography{tauris_refs}

\end{document}